\documentclass[a4paper]{jpconf}
\usepackage{graphicx}
\usepackage{hyperref}
\usepackage[capitalize]{cleveref}
\usepackage[utf8]{inputenc}

\begin{document}

\title{Dilepton production with the SMASH model}
\author{Janus Weil${}^1$, Jan Staudenmaier${}^{1,2}$, Hannah Petersen${}^{1,2,3}$}
\address{${}^1$ Frankfurt Institute for Advanced Studies, 60438 Frankfurt, Germany}
\address{${}^2$ Institut für Theoretische Physik, Goethe-Universität, 60438 Frankfurt, Germany}
\address{${}^3$ GSI Helmholtzzentrum für Schwerionenforschung GmbH, 64291 Darmstadt, Germany}
\ead{weil@fias.uni-frankfurt.de}

\begin{abstract}
In this work the SMASH model is presented (``Simulating Many Accelerated Strongly-Interacting Hadrons''),
a next-generation hadronic transport approach, which is designed to describe the
non-equilibrium evolution of hadronic matter in heavy-ion collisions.
We discuss first dilepton spectra obtained with SMASH in the few-GeV energy range
of GSI/FAIR, where the dynamics of hadronic matter is dominated by the
production and decay of various resonance states.
In particular we show how electromagnetic transition form factors can emerge in
a transport picture under the hypothesis of vector-meson dominance.
\end{abstract}

\section{Introduction}

Lepton pairs are a useful probe for the regions of high density and temperature
that are produced in a heavy-ion collision. They have been measured
experimentally by detectors like NA60 \cite{Arnaldi:2009aa} and HADES
\cite{HADES:2011ab}. In this work, the production of dileptons is investigated
within a hadronic transport approach, SMASH, which essentially solves the
Boltzmann equation for a hadron resonance gas. Here we focus on a few aspects of
the model that are important for dilepton production. A full description of
SMASH will be provided in an upcoming publication \cite{smash_overview}.

At low energies (of a few GeV) the particle interactions in SMASH rely on cross
sections that are dominated by the excitation and decay of hadronic resonance
states. A typical dilepton spectrum contains contributions from direct and
Dalitz decays of resonances such as:
\begin{itemize}
 \item $\rho,\omega,\phi\to e^+e^-$
 \item $\pi^0,\eta,\eta'\to e^+e^-\gamma$
 \item $\omega,\phi\to e^+e^-\pi$
 \item $N^*,\Delta\to e^+e^-N$
\end{itemize}
Here we mainly focus on the contributions from the $\omega$ meson. The major
production channels of the $\omega$ in pp and AA collisions at low energies are
assumed to be processes like $NN\to NN^*\to NN\omega$ or
$NN\to\Delta N^*\to NN\pi\omega$, where the $N^*\to\omega N$ couplings for the
$N^*$ states are taken from the PDG database \cite{Agashe:2014kda}.
The dominant $\omega$ decay channel is $\omega\to3\pi$, which in SMASH is
treated via the two-step decay chain $\omega\to\pi\rho\to3\pi$, the so-called
GSW process (Gell-Mann, Sharp, Wagner) \cite{GellMann:1962jt}. This is done in
order to be able to fulfill detailed balance with $1\leftrightarrow2$ and
$2\leftrightarrow2$ processes only (avoiding the technically more challenging
$1\leftrightarrow3$ processes). This assumption for the $\omega$ meson is also
used in other related works \cite{Muhlich:2006ps,Santini:2008pk}. Further it
should be noted that SMASH uses the decay-width parametrizations of
Manley et al.~\cite{Manley:1992yb}, which are also employed in the GiBUU
transport model \cite{Buss:2011mx}. The above assumptions lead to the fact that
in SMASH the $\omega$ meson is exclusively produced via the decay of $N^*$
resonances in pp collisions, while in AA collisions also the production via
$\pi\rho$ and $\pi\pi$ collisions is possible.

In the following we first investigate the $\omega$ Dalitz decay into
$e^+e^-\pi^0$ (which proceeds as a two-step chain via an intermediate $\rho$)
and then the Dalitz decay $N^*\to e^+e^-N$, which can proceed via an
intermediate $\rho$ or $\omega$ meson and is also treated as a two-step process.

\section{$\omega$ Dalitz decay}

We investigate the electromagnetic Dalitz decay of the $\omega$ meson,
$\omega\to\pi^0e^+e^-$, under the assumption that this decay proceeds in two
steps as $\omega\to\pi^0\rho^0\to\pi^0e^+e^-$. As mentioned above, this
assumption is motivated mainly from the fact that the SMASH model currently
emulates the $\omega\to3\pi$ decay by $\omega\to\pi\rho$, in order to fulfill
detailed balance without the need of introducing three-body collisions.
Furthermore, this assumption is in fact equivalent to the two-step VMD treatment
that was employed in earlier works for the electromagnetic decays of $N^*$ and
$\Delta^*$ baryons \cite{Weil:2012ji} and later extended also to the
$\Delta(1232)$ \cite{Weil:2014lma}.

\begin{figure}[h]
 \includegraphics[width=0.5\textwidth]{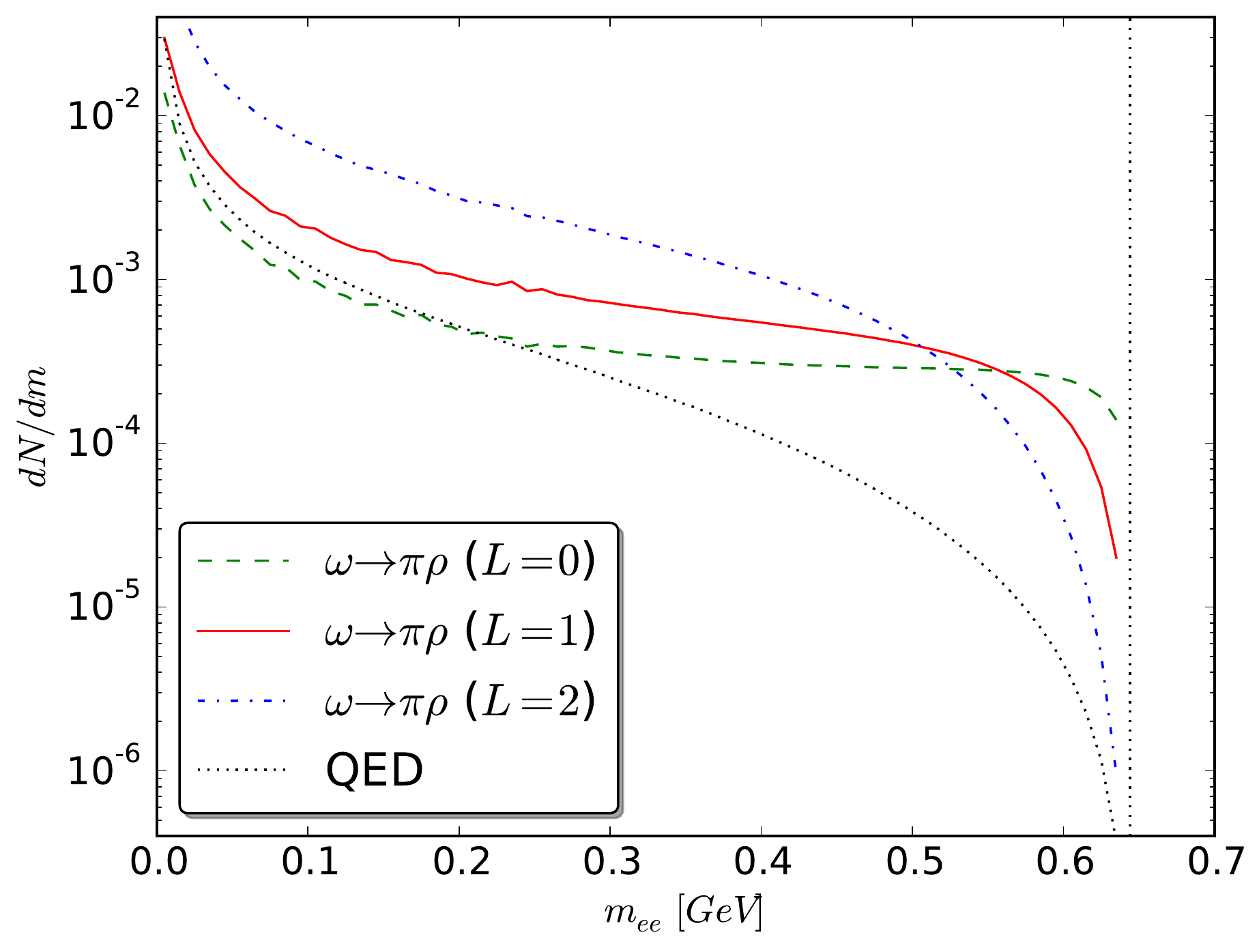}
 \includegraphics[width=0.5\textwidth]{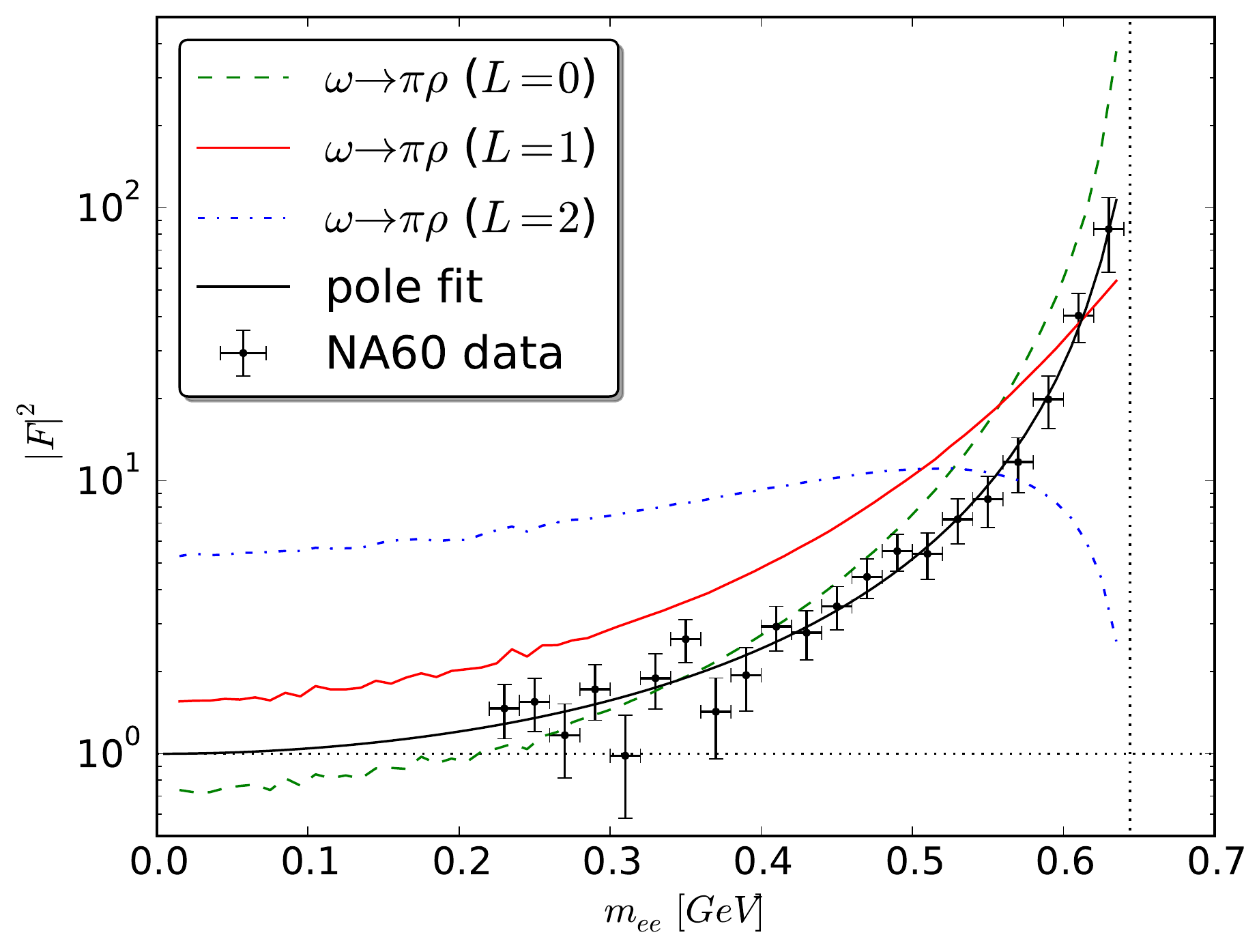}
 \caption{\label{fig1} $\omega\to\pi^0e^+e^-$ decay spectrum (left) and form factor (right)
          for a branching ratio of 89\% for $\omega\to\pi\rho$ and different angular momenta.
          The form factor is compared to data from the NA60 experiment \cite{Arnaldi:2009aa}.}
\end{figure}

In \cref{fig1} (left) we show the dilepton mass spectrum originating from the
process $\omega\to\pi\rho\to\pi e^+e^-$, using three different values for the
angular momentum $L$ in the $\omega\to\pi\rho$ decay and a branching ratio of
$\mathrm{BR}(\omega\to\pi\rho)=89\%$, such as to fully saturate the $3\pi$ decay.
The  values of $L=0,1,2$ are the only ones allowed by angular momentum
conservation, but in fact only $L=1$ is allowed by parity conservation, which
demands that $L$ is odd in order to fulfill $P_\omega = P_\pi \cdot P_\rho \cdot (-1)^L$,
where all three mesons have negative intrinsic parity. We actually show all
three cases in order to illustrate the effect of the angular momentum. The plot
also shows the QED case, which corresponds to a point-like vertex in the
$\omega$ Dalitz decay \cite{Landsberg:1986fd}. All curves assume that the
decaying $\omega$ meson is on the pole mass of $m_\omega=782$ MeV. The vertical
dotted line indicates the maximum dilepton mass of $m_\omega-m_\pi\approx 644$ MeV.

\begin{figure}[h]
 \centering
  \includegraphics[width=0.5\textwidth]{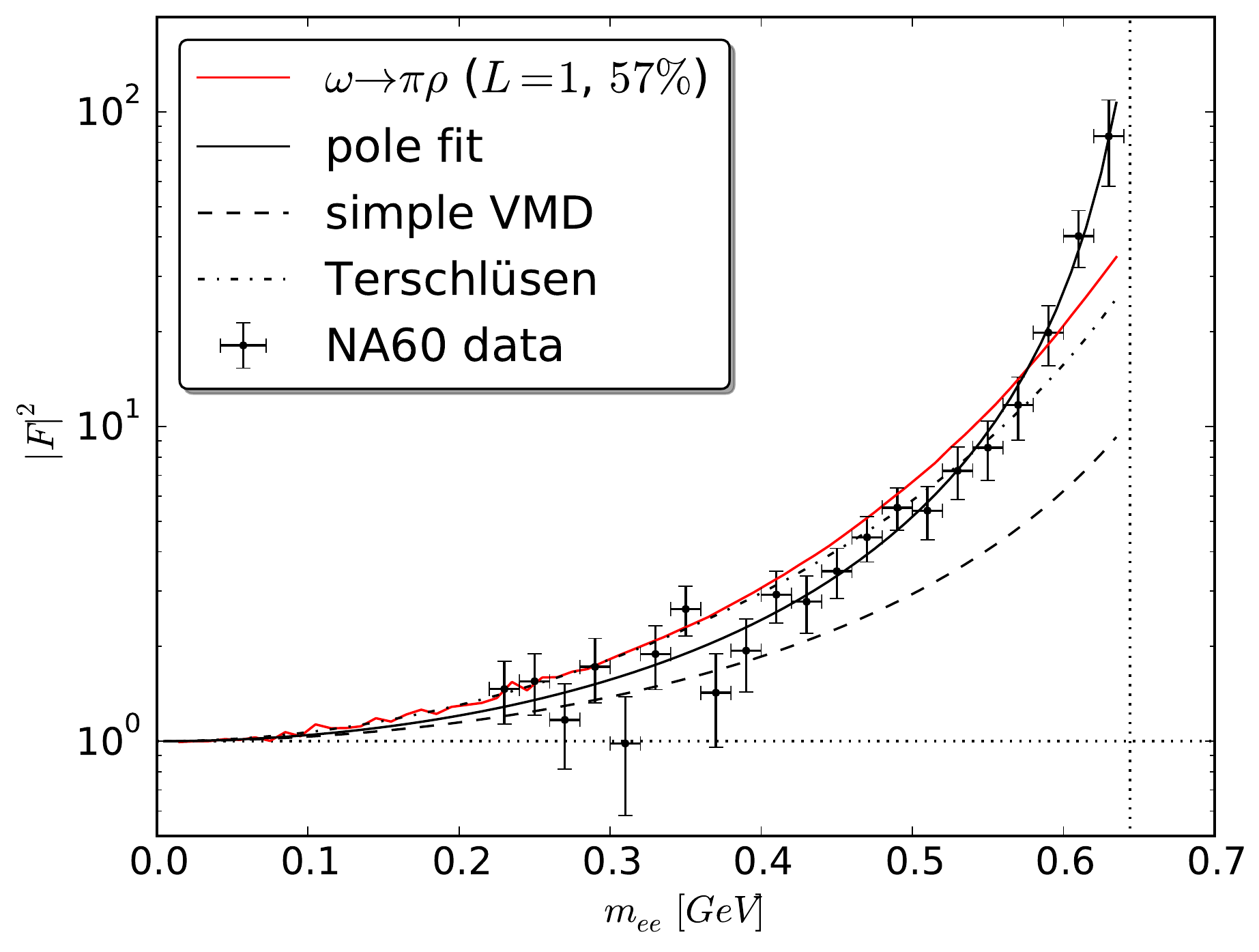}
 \caption{\label{fig2} $\omega\to\pi^0e^+e^-$  transition form factor with reduced $\omega\to\pi\rho$ (p-wave) coupling.}
\end{figure}

The right panel of \cref{fig1} shows the form factor of the $\omega$ Dalitz
decay (which is obtained simply by dividing the ``two-step VMD'' curves by the
QED baseline) compared to the data taken by NA60 \cite{Arnaldi:2009aa} and a fit
of the data using the pole approximation
\begin{equation}
 |F|^2 = (1-m^2/\Lambda^2)^{-2},
\end{equation}
with $\Lambda=668$ MeV. We note that the form factor is by definition normalized
to one at the real-photon point ($m=0$). However, this is not given a priori for
our three cases. Instead their normalization is determined by the branching
ratio, which we fix to 89\% for now (i.e. the upper limit given by the $3\pi$
decay). While the $L=2$ case is apparently in strong conflict with the data,
$L=0$ seems to be the best match for the data from a phenomenological point of
view. However, it is physically forbidden, since parity conservation only allows
$L=1$. The curve for $L=1$ touches the data only at the highest masses, while it
overshoots the data over most of the spectrum and also overestimates the
photon-point value. This is obviously due to the large branching ratio employed
for $\omega\to\pi\rho$. From comparing the $L=1$ form factor to the NA60 data,
one has to conclude that the assumption that the whole $\omega\to3\pi$ decay
proceeds via an intermediate $\pi\rho$ state was too strong. Therefore, we
reduce the $\omega\to\pi\rho$ branching, in order to avoid overshooting the form
factor data. In particular we scale it down from 89\% to 57\% to obtain the
right photon-point normalization. As shown in \cref{fig2}, this results in a
reasonable description of the experimental data. A significant deviation is
visible only for the two highest data points.

For comparison we also show a 'simple' VMD curve, as presented in the NA60 paper
\cite{Arnaldi:2009aa}, which only includes a plain $\rho$ propagator in pole
approximation.
Our treatment of $\omega\to\pi\rho\to\pi e^+e^-$ is of course also a VMD-like
model, but in contrast to the 'simple' VMD it includes a hadronic form factor
in the $\omega$-$\pi$-$\rho$ vertex.

In our model the electromagnetic transition form factor of the $\omega$ meson is
essentially determined by the mass distribution of the $\rho$ meson in the
$\omega\to\pi\rho$ decay, which is given by the following formula (leaving out
constant factors):
\begin{equation} \label{eq2}
 \frac{d\Gamma_{\omega\to\pi\rho}}{dm_\rho} \propto \mathcal{A}_\rho(m_\rho) \cdot p_F(m_\omega, m_\rho, m_\pi) \cdot B_L^2(p_FR)
\end{equation}
Following from eq.~(172)-(173) in \cite{Buss:2011mx}, this equation illustrates
that there are three factors that influence the mass distribution of the $\rho$
meson in the decay of the $\omega$:
\begin{enumerate}
 \item the spectral function $\mathcal{A}_\rho$ of the $\rho$ meson, which we
       assume to be a Breit-Wigner function with mass-dependent width
       \cite{Weil:2013mya},
 \item the phase-space factor $p_F$, i.e. the center-of-mass momentum in the
        $\pi$-$\rho$ final state and
 \item the Blatt-Weisskopf factor $B_L$, which depends on the angular momentum
       $L$ and includes a cutoff parameter $R$. For $L=1$,
       $B_1^2(x)=x^2/(1+x^2)$ is obtained.
\end{enumerate}
The Blatt-Weisskopf function contains both a phase-space factor as well as a
hadronic form factor. The latter is characterized by the cutoff parameter $R$,
which is taken to be $R=1$ fm.

Our model suggests that the $\omega$ meson decays into $\pi\rho$ with a
branching ratio of around 57\%, which makes up more than half of the $3\pi$
final state and at the same time provides a rather good explanation of the
electromagnetic transition form factor of the $\omega$ meson.
We also show in \cref{fig2} the result of the effective-Lagrangian approach of
Terschlüsen et al.~\cite{Terschlusen:2013iqa}, which in fact is rather close to
our result. Further we note that preliminary calculations seem to indicate that
our approach can also be applied to the $\phi\to e^+e^-\pi^0$ decay and yields
results compatible with recent data from the KLOE-2 collaboration
\cite{Anastasi:2016bfh}.

\section{Baryonic Dalitz decays}

Next we discuss the baryonic Dalitz decays, using as an example the process
$pp\to e^+e^-X$ at $E_{kin}=3.5$ GeV, as measured by the HADES collaboration
\cite{HADES:2011ab}. Our simulations do not yet include a detector-acceptance
filtering, therefore we can not directly compare to data but only want to
discuss qualitative effects here.

\begin{figure}[h]
 \includegraphics[width=0.5\textwidth]{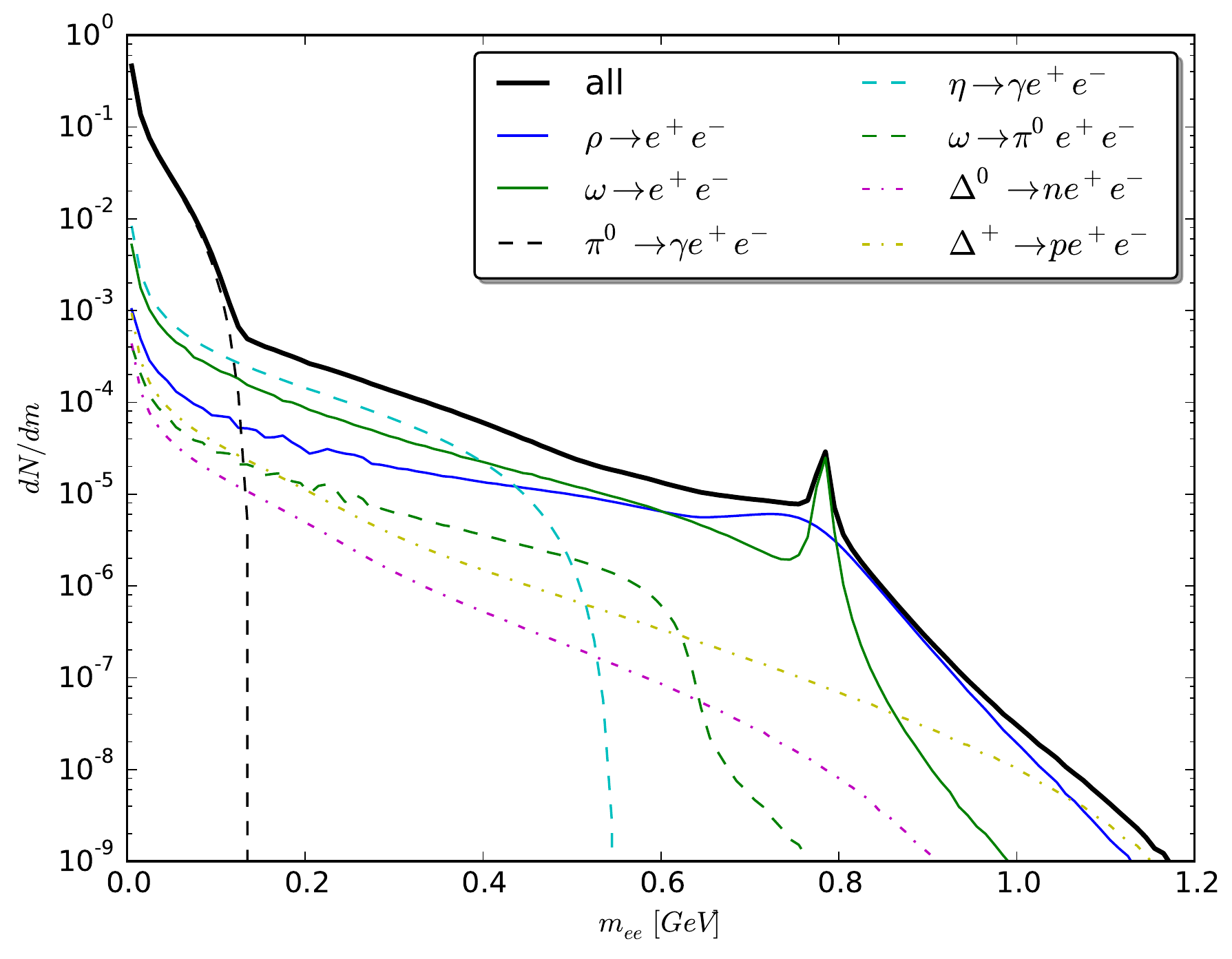}
 \includegraphics[width=0.5\textwidth]{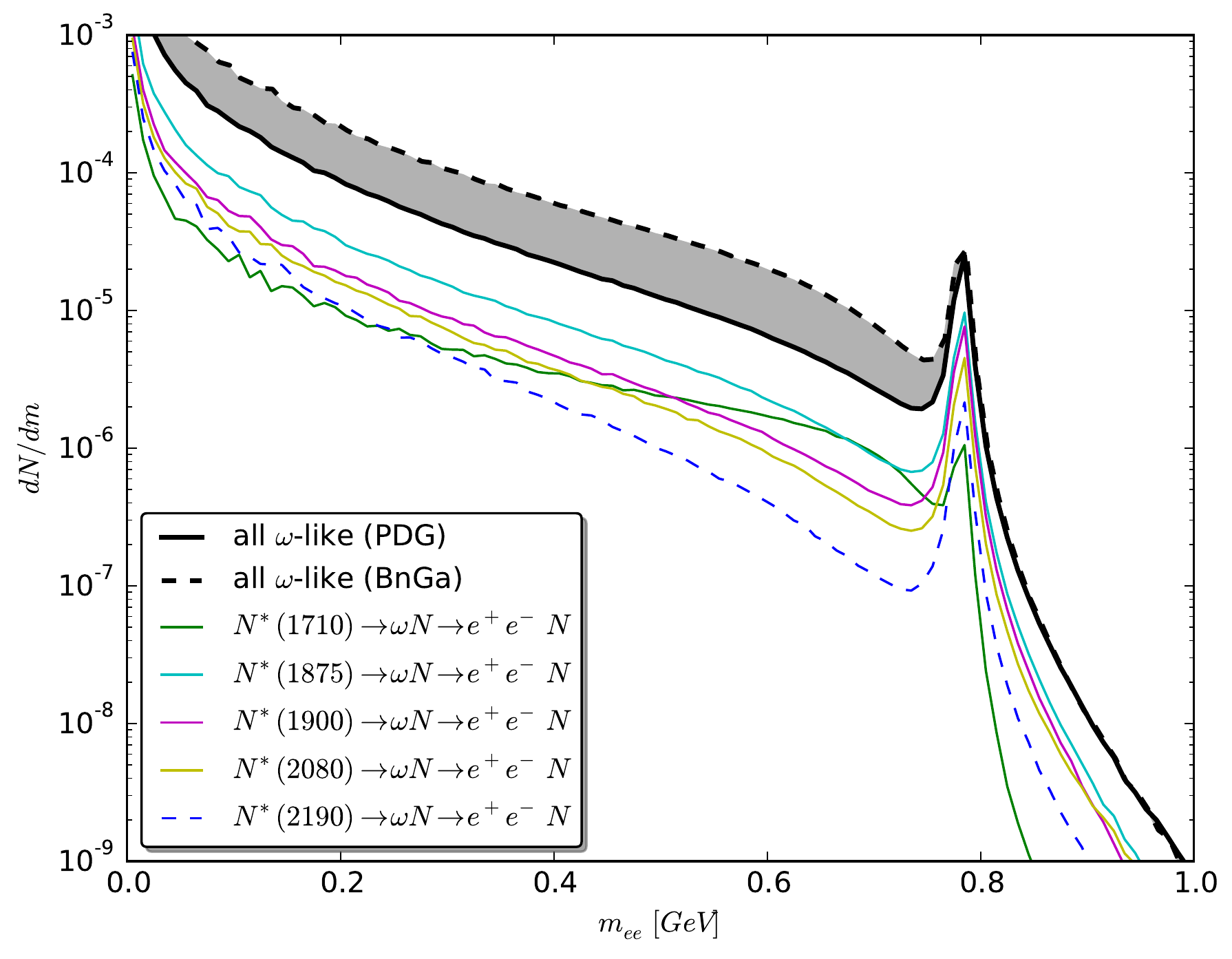}
 \caption{\label{fig3}Dilepton mass spectra for p+p at 3.5 GeV. Left: Total inclusive spectrum.
          Right: $\omega\to e^+e^-$ only, with contributions from different baryon resonances.}
\end{figure}

\Cref{fig3} (left) shows the inclusive dilepton mass spectrum simulated with the
SMASH model. We note that the $\pi^0$, $\eta$ and $\Delta$ Dalitz decays follow
the standard treatment employed in many other transport approaches (see
e.g.~\cite{Weil:2013mya}), the first two with a parametrized form factor, the
latter in QED approximation. The $\rho\rightarrow e^+e^-$ contribution consists
mainly of $N^*\to\rho N\to e^+e^-N$ and $\Delta^*\to\rho N\to e^+e^-N$ processes,
since almost all $\rho$ mesons in pp collisions are produced via baryonic
resonance decays in SMASH. This is why the $\rho$ contribution has a Dalitz-like
shape with contributions below the $2\pi$ threshold, which has already been seen
in earlier GiBUU simulations \cite{Weil:2012ji}. In addition to the baryonic
Dalitz decays, the $\rho$ contains an additional feed-down from the
$\omega\to\pi\rho$ decay, as discussed in the previous section.
The most interesting feature of the SMASH dilepton spectrum is actually the
$\omega\to e^+e^-$ contribution. In addition to the peak at the $\omega$ pole
mass of 782 MeV, it also contains a strong Dalitz-like tail in the low-mass
region. This is due to the fact that the $\omega$ in SMASH is produced mainly
via the excitation and decay of baryonic resonances in pp collisions, just like
the $\rho$ meson. Due to isospin arguments, the production of the $\omega$
proceeds only via $N^*$ resonances, while the $\rho$ can also be produced from
$\Delta^*$ decays.

The right panel of \cref{fig3} shows the $\omega$ contribution to the dilepton
spectrum, with all the contributions from different $N^*$ states. The branching
ratios for $N^*\to\omega N$ are taken from the current PDG database by default
\cite{Agashe:2014kda}. We also show how the spectrum changes when using the
branching ratios from a recent analysis of the Bonn-Gatchina group
\cite{Denisenko:2016ugz}. This analysis yields strong $\omega$ couplings for
sub-threshold resonances like the $N^*(1700)$ and $N^*(1720)$ that do not have
an $\omega N$ mode in the current PDG, which further enhances the low-mass
Dalitz tail (while the yield at the $\omega$ pole mass and above is basically
unchanged).
Due to the missing acceptance filtering, we can not yet comment whether these
effects are compatible with experimental data. The general fact that effects
from baryonic Dalitz decays exist not only for the $\rho$, but also for the
$\omega$ meson is not extremely surprising. However, the sheer size of the
effect is rather large in our preliminary simulations, and certainly requires
further investigation.

\section{Conclusions}

We have shown that the first simulations of dilepton spectra with the SMASH
model yield some interesting effects concerning the $\omega$ meson. Firstly, our
model for the $\omega\to\pi\rho$ p-wave decay shows a rather good agreement with
NA60 data for the electromagnetic $\omega$ transition form factor and suggests a branching
ratio of roughly 57\% for the GSW process $\omega\to\pi\rho$. Secondly we have
investigated effects of the baryonic coupling of the $\omega$ meson, which can
have an impact on dilepton spectra from pp as well as AA collisions. Our
preliminary results indicate that such effects could be surprisingly large, but
need further investigation before final conclusions can be drawn.

\section*{Acknowledgements}

The authors thank S.~Leupold and P.~Salabura for useful comments and
S.~Damjanovic for providing the NA60 data points.
J.W. and H.P. acknowledge funding of a Helmholtz Young Investigator Group
VH-NG-822 from the Helmholtz Association and GSI.
This work was supported by the Helmholtz International Center for the Facility
for Antiproton and Ion Research (HIC for FAIR) within the framework of the
Landes-Offensive zur Entwicklung Wissenschaftlich-Oekonomischer Exzellenz
(LOEWE) program launched by the State of Hesse.
Computational resources have been provided by the Center for Scientific
Computing (CSC) at the Goethe-University of Frankfurt.

\section*{References}

\bibliographystyle{iopart-num}
\bibliography{inspire,others}

\providecommand{\newblock}{}
\begin{thebibliography}{10}
\expandafter\ifx\csname url\endcsname\relax
  \def\url#1{{\tt #1}}\fi
\expandafter\ifx\csname urlprefix\endcsname\relax\def\urlprefix{URL }\fi
\providecommand{\eprint}[2][]{\url{#2}}

\bibitem{Arnaldi:2009aa}
Arnaldi R {\em et~al.\/} (NA60) 2009 {\em Phys. Lett. B\/} {\bf 677} 260--266
  (\textit{Preprint} \eprint{0902.2547})

\bibitem{HADES:2011ab}
Agakishiev G {\em et~al.\/} (HADES) 2012 {\em Eur. Phys. J. A\/} {\bf 48} 64
  (\textit{Preprint} \eprint{1112.3607})

\bibitem{smash_overview}
Weil J {\em et~al.\/} {\em (in preparation)\/}

\bibitem{Agashe:2014kda}
Olive K~A {\em et~al.\/} (Particle Data Group) 2014 {\em Chin. Phys. C\/} {\bf
  38} 090001

\bibitem{GellMann:1962jt}
Gell-Mann M, Sharp D and Wagner W~G 1962 {\em Phys. Rev. Lett.\/} {\bf 8} 261

\bibitem{Muhlich:2006ps}
Muhlich P and Mosel U 2006 {\em Nucl. Phys. A\/} {\bf 773} 156--172
  (\textit{Preprint} \eprint{nucl-th/0602054})

\bibitem{Santini:2008pk}
Santini E, Cozma M~D, Faessler A, Fuchs C, Krivoruchenko M~I and Martemyanov B
  2008 {\em Phys. Rev. C\/} {\bf 78} 034910 (\textit{Preprint}
  \eprint{0804.3702})

\bibitem{Manley:1992yb}
Manley D~M and Saleski E~M 1992 {\em Phys. Rev. D\/} {\bf 45} 4002--4033

\bibitem{Buss:2011mx}
Buss O, Gaitanos T, Gallmeister K, van Hees H, Kaskulov M, Lalakulich O,
  Larionov A~B, Leitner T, Weil J and Mosel U 2012 {\em Phys. Rept.\/} {\bf
  512} 1--124 (\textit{Preprint} \eprint{1106.1344})

\bibitem{Weil:2012ji}
Weil J, van Hees H and Mosel U 2012 {\em Eur. Phys. J. A\/} {\bf 48} 111
  (\textit{Preprint} \eprint{1203.3557})

\bibitem{Weil:2014lma}
Weil J, Endres S, van Hees H, Bleicher M and Mosel U 2014 {\em {Proceedings,
  20th International Conference on Particles and Nuclei (PANIC 14)}\/} pp
  290--293 (\textit{Preprint} \eprint{1410.4206})

\bibitem{Landsberg:1986fd}
Landsberg L~G 1985 {\em Phys. Rept.\/} {\bf 128} 301--376

\bibitem{Weil:2013mya}
Weil J 2013 {\em {Vector Mesons in Medium in a Transport Approach}\/} Ph.D.
  thesis Giessen U.
  \urlprefix\url{http://geb.uni-giessen.de/geb/volltexte/2013/10253/}

\bibitem{Terschlusen:2013iqa}
Terschlüsen C, Strandberg B, Leupold S and Eichstädt F 2013 {\em Eur. Phys.
  J. A\/} {\bf 49} 116 (\textit{Preprint} \eprint{1305.1181})

\bibitem{Anastasi:2016bfh}
Anastasi A {\em et~al.\/} (KLOE-2) 2016 {\em Phys. Lett. B\/} {\bf 757}
  362--367 (\textit{Preprint} \eprint{1601.06565})

\bibitem{Denisenko:2016ugz}
Denisenko I, Anisovich A~V, Crede V, Eberhardt H, Klempt E, Nikonov V~A,
  Sarantsev A~V, Schmieden H, Thoma U and Wilson A 2016 {\em Phys. Lett. B\/}
  {\bf 755} 97--101 (\textit{Preprint} \eprint{1601.06092})

\end{thebibliography}

\end{document}